\documentclass[a4paper]{jpconf}
\usepackage{graphicx}
\usepackage{subfigure}

\begin{document}

\title{Measurement of hadron suppression and study of its connection with vanishing $v_{\mathrm{3}}$ at low $\sqrt{s_{\mathrm{NN}}}$ in Au+Au collisions with STAR}

\author{Stephen Horvat (for the STAR Collaboration)}

\address{Yale University, New Haven, CT, USA}

\begin{abstract}
At top RHIC and LHC energies the suppression of high transverse momentum ($p_{\mathrm{T}}$) hadrons provides evidence for partonic energy loss in QGP. We study partonic energy loss in the RHIC Beam Energy Scan (BES) by investigating the centrality dependence of the binary-collision-scaled high-$p_{\mathrm{T}}$ yields.  Observing a decrease of the scaled yield in more central collisions is proposed as a possible signature for jet quenching.   Even at energies and centralities where this signature is lost a QGP may still be formed since competing  phenomena responsible for enhancements may overwhelm the suppression from energy loss.  Measurements in several ranges of $p_{\mathrm{T}}$ from $\sqrt{s_{_{\mathrm{NN}}}}$ = 7.7, 11.5, 14.5, 19.6, 27, 39, 62.4, and 200 GeV data show that relative hadron suppression persists at least down to 14.5 GeV. 

To further investigate both the possible formation of a QGP at these lower energies and whether the observed hadron suppression coincides with the onset of other QGP signatures, we examine the energy and centrality dependence of $v_{3}$. Models have shown that the development of $v_{3}$ requires the presence of a low viscosity phase early in the collision. We find that for collisions with $N_{\mathrm{part}}$\textless50, $v_{3}$ disappears for energies below 14.5 GeV, suggestive of a turn-off of the QGP. But for $N_{\mathrm{part}}$\textgreater50, $v_{3}$  persists down to the lowest energies.  Together, these signatures provide possible evidence for the formation of a QGP in the lowest energy collisions at RHIC.

\end{abstract}

\section{Introduction}
\label{}
\begin{figure}
\begin{center}
\includegraphics[width=20pc]{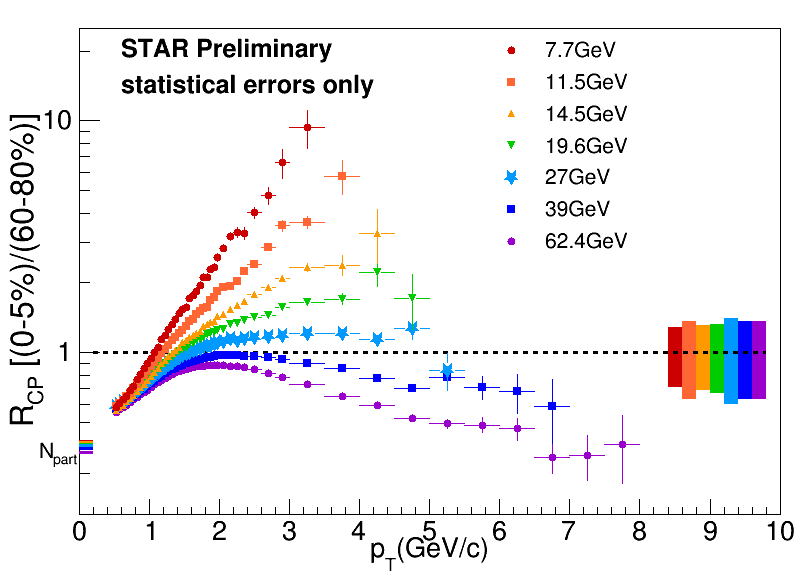}
\end{center}
\caption{\label{labela}Charged hadron $R_{\mathrm{CP}}$ for RHIC BES energies.  The error bands at unity on the right side of the plot correspond to the $p_{\mathrm{T}}$ independent systematic uncertainty in $N_{\mathrm{bin}}$ scaling with the color in the band corresponding to the color of the data points for that energy.  The vertical error bars correspond to statistical errors.}
\end{figure}
The strong force binds, or confines, quarks and gluons into color neutral hadrons.  By colliding heavy ions at ultra-relativistic energies, local energy densities and temperatures increase and a new state of deconfined quarks and gluons (QGP) may be produced.  Several signatures for the formation of QGP have been proposed and measured \cite{Arsene:2004fa,Back:2004je,Adams:2005dq,Adcox:2004mh}.   For sufficiently low collision energies these signatures would be expected to turn off, with different signatures vanishing at different collision energies depending on how well initial state and hadronic phase effects can be disentangled.  The initial state effects include the Cronin Effect \cite{Cronin:1974zm,Antreasyan:1978cw,Straub:1992xd} and nuclear modifications to the parton distribution function (nPDF) \cite{Helenius:2012ny}.  One possible contribution from the nPDF would be the impact parameter dependence of the EMC effect \cite{Aubert:1983xm} which may contribute to suppression for some collision energies and momenta.  Smaller impact parameter collisions are more likely to produce a deconfined medium due to the larger overlap region and higher initial energy density.  These smaller impact parameters should also lead to more initial scattering and re-scattering (Cronin Effect), larger effects from the nPDF, and more final state particles and radial flow in the hadronic phase.  Besides partonic energy loss \cite{Bjorken:1982tu,Wang:1991xy}, impact parameter dependent nPDFs in the initial state and hadronic energy loss may also contribute to suppression.  

These effects leading to suppression must compete with enhancement from the Cronin Effect and radial flow.  The nuclear modification factor, $R_{\mathrm{CP}}$, compares the $N_{\mathrm{coll}}$-scaled $p_{\mathrm{T}}$-spectra in central, small impact parameter, collisions relative to those in peripheral collisions.  Spectra are modified by all of the above effects so that only a net enhancement or suppression is measured depending on which effects dominate.  Measurements of $R_{\mathrm{CP}}$ ($\sqrt{s_{_{\mathrm{NN}}}}$,$p_{\mathrm{T}}$), using 60-80\% spectra for a reference, find a strong enhancement of high-$p_{\mathrm{T}}$ charged hadrons for the lower collision energies (Fig. \ref{labela}).  If there are still strong enhancement effects at the energy where $R_{\mathrm{CP}}$ crosses unity then there must also be strong suppression effects at that energy.  This motivates the measurements highlighted here: looking at the centrality dependence of $R_{\mathrm{CP}}$ and looking for where $v_{\mathrm{3}}$ vanishes.  The centrality dependence of $R_{\mathrm{CP}}$ may help us to disentangle some of these effects if enhancement and suppression terms have different centrality dependences.  Since the most peripheral spectra are used as a reference for every centrality, it can be multiplied out and we can simply look at the centrality dependence of the $N_{\mathrm{coll}}$-scaled yields (Eq. \ref{eq:Y}),  \begin{equation} \label{eq:Y}
 Y(\langle N_{\mathrm{part}}\rangle ) =  \frac{\mathrm{1}}{\langle N_{\mathrm{coll}}\rangle }\frac{d^2N}{dp_{\mathrm{T}} d\eta}(\langle N_{\mathrm{part}}\rangle ). 
\end{equation}The $Y$ distributions are normalized by the yields in their most peripheral bins.  If the values of $Y$ in central collisions are lower than unity then $Y$ is suppressed relative to peripheral collisions.  If they are lower than any less central bin then they are suppressed relative to that bin.  

In addition to these high-$p_{\mathrm{T}}$ probes, another proposed signature for the turn-off of the QGP is the disappearance of $v_{\mathrm{3}}$; a measurement dominated by low-$p_{\mathrm{T}}$ particles \cite{PhysRevC.88.064908}.  Following a Fourier decomposition of two-particle correlations, the higher order harmonics are more sensitive to fine structure and are more suppressed by viscosity \cite{PhysRevC.91.024908}.  A hybrid model of UrQMD which includes a hydrodynamic phase has shown that a low viscosity medium is needed early in the collision in order to generate finite values of $v_{\mathrm{3}}$ \cite{PhysRevC.88.064908}.  

\section{Data and Results}
\label{}

\begin{figure}
\begin{center}
\includegraphics[width=30pc]{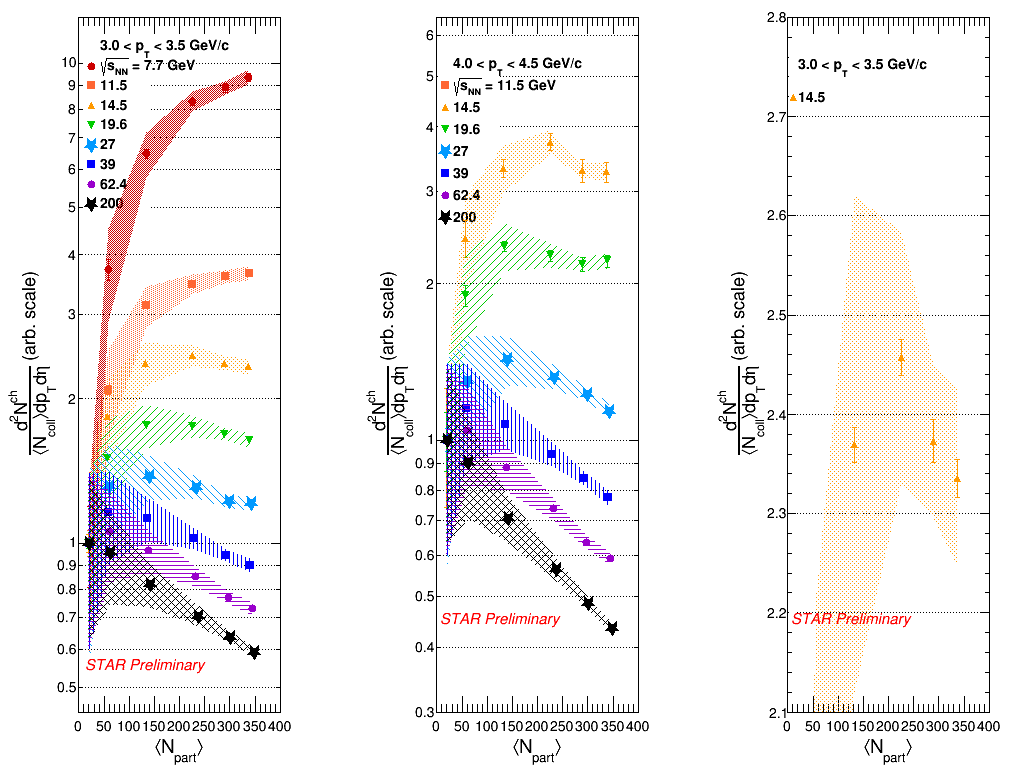}
\end{center}
\caption{\label{labeld}$Y(\langle N_{\mathrm{part}}\rangle$) in the RHIC BES for (left) 3.0\textless$p_{\mathrm{T}}$\textless3.5 GeV/c, (center) 4.0\textless$p_{\mathrm{T}}$\textless4.5 GeV/c and (right) zooming in on the 14.5 GeV distribution for 3.0\textless$p_{\mathrm{T}}$\textless3.5 GeV/c.  The error bands are $N_{ \mathrm{part}}$ scaling uncertainties and are completely correlated within each dataset.}
\end{figure}

The data were taken in phase one of the RHIC BES  \cite{Aggarwal:2010cw}.  The centrality, $\langle N_{\mathrm{part}}\rangle$ and $\langle N_{\mathrm{coll}}\rangle$ are determined from a Monte Carlo Glauber simulation \cite{Miller:2007ri}.  The uncertainties in $\langle N_{\mathrm{part}}\rangle$ and $\langle N_{\mathrm{coll}}\rangle$ are completely correlated as a function of centrality within each dataset and increase for more peripheral collisions, mostly due to the uncertainties in the trigger and vertex reconstruction efficiencies.  Additionally, the data were corrected for the Time Projection Chamber's tracking efficiency by using tracks from a \textsc{geant3} simulation of the STAR detector that are embedded into real events \cite{Brun:1994aa}.  

Charged hadron $R_{\mathrm{CP}}$ (Fig. \ref{labela}) is measured at mid-rapidity for 14.5 GeV and found to fit smoothly into the energy dependence that was observed at other energies.  By measuring $Y(\langle N_{\mathrm{part}}\rangle)$, which is analogous to the centrality dependence of $R_{\mathrm{CP}}$, we find a smooth evolution as a function of the centrality and energy of high-$p_{\mathrm{T}}$ yields (Fig. \ref{labeld}).  The values of $Y$ decrease monotonically in the 200 GeV data, consistent with suppression effects becoming more dominant with increasing centrality.  As the collision energy decreases, there is a rise in values of $Y$ initially, which then turns over and decreases as the collisions become more central.  As the energy continues to decrease, the turn-over gets pushed to more and more central events until for 11.5 and 7.7 GeV there is a monotonic rise. This matches a scenario where the effects leading to enhancement turn on first as the collisions go from peripheral to central, but as the collisions become more central suppression effects begin to dominate.  The centrality where suppression effects increase at the same rate as enhancement effects, the maximum of the $Y$ distribution, moves to more central collisions as the energy is decreased; which is consistent with the picture that a less dense and shorter-lived QGP was formed.

\begin{figure}
\begin{center}
\includegraphics[width=18pc]{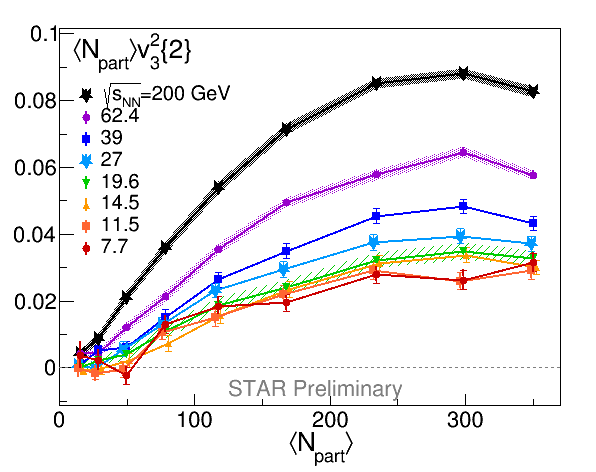}
\end{center}
\caption{\label{labelb}$\langle N_{\mathrm{part}}\rangle v_{\mathrm{3}}^{2}\{2\}$($\langle N_{\mathrm{part}}\rangle$) is shown for $\sqrt{s_{_{\mathrm{NN}}}}$ = 7.7, 11.5, 14.5, 19.6, 27, 39, 62.4, and 200 GeV.  With small energy dependence for the systematic uncertainties, representative systematic uncertainty bands are shown for three energies.}
\end{figure}

Triangular flow, $v_{\mathrm{3}}^{2}\{2\}$, has been measured for charged particles as a function of $\Delta\eta$.  The method used for measuring $v_{\mathrm{3}}^{2}\{2\}$ is described in \cite{Song:2015aa}.  Figure \ref{labelb} shows $\langle N_{\mathrm{part}}\rangle v_{\mathrm{3}}^{2}\{2\}$($\langle N_{\mathrm{part}}\rangle$) for BES energies.  A clear energy ordering is observed for the magnitude of $v_{\mathrm{3}}^{2}\{2\}$ in central collisions.  The values of $v_{\mathrm{3}}^{2}\{2\}$ are found to vanish for low energy peripheral collisions.  The lower the energy, the more central the collisions must be in order to generate a non-zero $v_{\mathrm{3}}^{2}\{2\}$.  

\section{Summary and Outlook}
Two possible signatures for the formation of a QGP have been investigated: $Y$ and $v_{\mathrm{3}}^{2}$\{2\}.  $Y$ exhibits suppression of charged particle high-$p_{\mathrm{T}}$ yields in central relative to less central collisions for$\sqrt{s_{_{\mathrm{NN}}}}$$\geq$14.5 GeV.  While consistent with expectations, caution must be used before concluding that this relative suppression of $Y$ is a result of partonic energy loss.  Alternative explanations, such as hadronic energy loss, must be ruled out first.  $p$+$p$ and $p$+$A$ reference data for the BES, such as the $d$+$Au$ collisions that RHIC will run at several energies in 2016, will help in quantifying the modifications due to initial state effects.  The $v_{3}$ measurements do not share this ambiguity since $v_{3}$ and higher harmonics provide sensitivity to the viscosity at early times \cite{PhysRevC.91.024908}.  STAR's measurements of $v_{\mathrm{3}}^{2}$\{2\}($\langle N_{\mathrm{part}}\rangle$) are consistent with the formation of a low viscosity medium for $N_{\mathrm{part}}$\textgreater 50 and $\sqrt{s_{_{\mathrm{NN}}}}$$\geq$7.7 GeV.  This means that both of these possible signatures are found to ``turn-off'' for sufficiently low energy peripheral collisions. 

\section{References}

\bibliography{signatures_arxiv}

\begin{thebibliography}{10}
\expandafter\ifx\csname url\endcsname\relax
  \def\url#1{\texttt{#1}}\fi
\expandafter\ifx\csname urlprefix\endcsname\relax\def\urlprefix{URL }\fi
\expandafter\ifx\csname href\endcsname\relax
  \def\href#1#2{#2} \def\path#1{#1}\fi

\bibitem{Arsene:2004fa}
I.~Arsene, et~al., Nucl.Phys. A \textbf{757} (2005) 1--27.

\bibitem{Back:2004je}
B.~Back, et~al., Nucl.Phys. A \textbf{757} (2005) 28--101.

\bibitem{Adams:2005dq}
J.~Adams, et~al., Nucl.Phys. A \textbf{757} (2005) 102--183.

\bibitem{Adcox:2004mh}
K.~Adcox, et~al., Nucl.Phys. A \textbf{757} (2005) 184--283.

\bibitem{Cronin:1974zm}
J.~Cronin, H.~J. Frisch, M.~Shochet, J.~Boymond, R.~Mermod, et~al., Phys.Rev. D
  \textbf{11} (1975) 3105.

\bibitem{Antreasyan:1978cw}
D.~Antreasyan, J.~Cronin, H.~J. Frisch, M.~Shochet, L.~Kluberg, et~al.,
  Phys.Rev. D \textbf{19} (1979) 764--778.

\bibitem{Straub:1992xd}
P.~Straub, D.~Jaffe, H.~D. Glass, M.~Adams, C.~Brown, et~al., Phys.Rev.Lett.
  \textbf{68} (1992) 452--455.

\bibitem{Helenius:2012ny}
I.~Helenius, K.~Eskola, H.~Honkanen, C.~Salgado, Nucl.Phys. A
  \textbf{904}-\textbf{905} (2013) 999c--1002c.

\bibitem{Aubert:1983xm}
J.~J. Aubert, et~al., Phys. Lett. B \textbf{123} (1983) 275.

\bibitem{Bjorken:1982tu}
J.~Bjorken, FERMILAB-PUB-82-059-THY.

\bibitem{Wang:1991xy}
X.-N. Wang, M.~Gyulassy, Phys.Rev.Lett. \textbf{68} (1992) 1480--1483.

\bibitem{PhysRevC.88.064908}
J.~Auvinen, H.~Petersen, Phys. Rev. \textbf{C} 88 (2013) 064908.

\bibitem{PhysRevC.91.024908}
C.~Shen, U.~Heinz, J.-F. m.~c. Paquet, I.~Kozlov, C.~Gale, Phys. Rev.
  \textbf{C} 91 (2015) 024908.

\bibitem{Aggarwal:2010cw}
M.~Aggarwal, et~al.\href {http://arxiv.org/abs/1007.2613}
  {\path{arXiv:1007.2613}}.

\bibitem{Miller:2007ri}
M.~L. Miller, K.~Reygers, S.~J. Sanders, P.~Steinberg, Ann.Rev.Nucl.Part.Sci.
  \textbf{57} (2007) 205--243.

\bibitem{Brun:1994aa}
R.~Brun, F.~Carminati, S.~Giani, {GEANT Detector Description and Simulation
  Tool}.

\bibitem{Song:2015aa}
{L. Song (STAR Collaboration)}, {Proceedings from this conference}.

\end{thebibliography}

\end{document}